\documentstyle[12pt]{article}
\newlength{\extraspace}
\newlength{\extraspaces}
\newcommand{\be}{\begin{equation}
\addtolength{\abovedisplayskip}{\extraspaces}
\addtolength{\belowdisplayskip}{\extraspaces}
\addtolength{\abovedisplayshortskip}{\extraspace}
\addtolength{\belowdisplayshortskip}{\extraspace}}
\newcommand{\ee}{\end{equation}}
\newcommand{\ba}{\begin{eqnarray}
\addtolength{\abovedisplayskip}{\extraspaces}
\addtolength{\belowdisplayskip}{\extraspaces}
\addtolength{\abovedisplayshortskip}{\extraspace}
\addtolength{\belowdisplayshortskip}{\extraspace}}
\newcommand{\nonu}{\nonumber \\[.5mm]}
\newcommand{\A}{&\!\!\!}
\newcommand{\ea}{\end{eqnarray}}
\newcommand{\case}[4]
{\left\{
\begin{array}{ll}
#1 & \mbox{ ( #2 ) } \\
#3 & \mbox{ ( #4 ) }
\end{array}
\right.}
\newcommand{\Pb}[2]{\left\{ #1 \: , #2 \right\}_{P.B.}}
\newcommand{\com}[2]{\left[ #1 \: , #2 \right]}
\newcommand{\acom}[2]{\left\{ #1 \: , #2 \right\}}
\newcommand{\iN}{\quad (i=1, \ldots ,N)}
\newcommand{\ijN}{\quad (i,j=1, \ldots ,N)}
\newcommand{\sumN}{\sum_{i=1}^N}
\newcommand{\sumz}[1]{\sum_{#1 \in {\bf Z}}}
\newcommand{\iu}{\sqrt{-1} \,}
\newcommand{\pa}[1]{\partial_{#1}}
\newcommand{\e}[1]{\, {\rm e}^{#1}}
\newcommand{\slr}{SL(2,{\bf R})}
\newcommand{\slc}{SL(2,{\bf C})}
\newcommand{\nm}[1]{\,: #1 : \,}
\newcommand{\hc}[1]{{#1}^{\dagger}}
\newcommand{\bra}[1]{\left\langle \, {#1} \, \right\vert}
\newcommand{\ket}[1]{\left\vert \, {#1} \, \right\rangle}

\newcommand{\sumn}{\sum_{n \neq 0}}

\newcommand{\tvo}[2]{{V^{( \vec{p}_{ #1 } )} ( x^{1}_{ #2 } )}}

\newcommand{\w}[2]{{W^{( \vec{p}_{ #1 } )} ( x^{1}_{ #2 } )}}
\newcommand{\sel}[3]{{ #1 }^{\vec{p}_{ #2 } \cdot \vec{p}_{ #3 }}}
\newcommand{\G}[1]{\Gamma \left( #1 \right)}
\newcommand{\J}[3]{J_{ #1 , #2 , #3 } ( \alpha , \beta ; \rho )}
\begin{document}
\begin{titlepage}
\begin{flushright}
TIT/HEP--216 \\
\today
\end{flushright}
\vspace{10mm}
\begin{center}
{\Large\bf Canonical Theory of 2D Gravity \\[2mm]
Coupled to Conformal Matter} \\[25mm]
{\sc Taku Uchino} \\[3mm]
{\it Department of Physics, Tokyo Institute of Technology \\[2mm]
Oh-okayama, Meguro, Tokyo 152, Japan} \\[30mm]
{\bf Abstract}\\[5mm]
\parbox{13cm}{
A canonical quantization for two dimensional gravity models,
including a dilaton gravity model,
is performed in a way suitable for the light-cone gauge.
We extend the theory developed by Abdalla {\it et.al.}\cite{AM}
and obtain the correlation functions
by using the screening charges of the symmetry algebra.
It turns out that the role of the cosmological constant term
in the Liouville theory is played by the screening charge
of the symmetry algebra and the resulting theory looks like a chiral part of
the Liouville theory or a non-critical open string theory.
Moreover,
we show that the dilaton gravity theory
has a symmetry realized by the centrally extended Poincar\'{e} algebra
instead of the $\slr$ Kac-Moody algebra which is a symmetry of
an ordinary two dimensional gravity theory.
}
\end{center}
\end{titlepage}
\section{Introduction}
\label{Introduction}
Since Polyakov and his collaborators solved the two dimensional
quantum gravity theory in the continuum
approach\cite{KPZ},
the $\slr$ symmetry of the two dimensional quantum gravity theory
has played an important role.
Although this symmetry is transparent in the light-cone gauge,
remarkable progress has been made in the conformal gauge
based on the results of papers \cite{DDK}.
For example,
correlation functions have been calculated by the path integral formalism
in the conformal gauge \cite{FK}.
However,
it would be better to know what happens in the canonical formalism
in the light-cone gauge.
In this paper we will investigate a canonical formalism
of the two dimensional gravity theory including the dilaton gravity theory
in a gauge independent way\footnote{
But this method is suitable for the light-cone gauge.}
which has been developed by Abdalla {\it et.al.} \cite{AM}.
We will show that in the case of the dilaton gravity theory \cite{CGHS},
the residual symmetry is represented
by the Kac-Moody version of the centrally extended Poincar\'{e} algebra
instead of the $\slr \otimes U(1)$ Kac-Moody algebra.
The action we will investigate is
\be
S = - \int_{\Sigma} d^2 x \sqrt{-det(g)}
\left(
  \frac{\alpha}{2} g^{\mu \nu} \pa{\mu} \phi \pa{\nu} \phi
  + \beta \phi R[g] + \Lambda_0
  + \frac{1}{2} \sumN
  g^{\mu \nu} \pa{\mu} \phi^i \pa{\nu} \phi^i
\right),
\label{our action}
\ee
where $\phi$ and $\phi^i$ $(i=1,\ldots,N)$ are scalar fields,
$g$ is the metric and $R[g]$ is the scalar curvature constructed from $g$.
$\Sigma$ is the space-time on which these fields are defined.
For simplicity,
we assume the space-time $\Sigma$ is a cylinder
with the coordinate system $(x^0,x^1)$,
where $x^0$ is a time coordinate which takes any real value,
while $x^1$ is a spatial one with values lying between $-\pi$ and $\pi$.
The action~(\ref{our action}) contains three parameters,
$\alpha$,
$\beta$ and $\Lambda_0$,
where $\Lambda_0$ is a cosmological constant.
We assume $\beta \neq 0$ in this paper.
The case $\alpha = 0$ corresponds to the two dimensional
dilaton gravity theory,
because the CGHS action \cite{CGHS}
\be
S_{CGHS} =
\int d^2 x \sqrt{-det(\overline{g})}
\left[
  \e{-2 \Phi} \left(
    R[\overline{g}] + 4 \overline{g}^{\mu \nu} \pa{\mu} \Phi \pa{\nu} \Phi
    - \Lambda_0
  \right)
- \frac{1}{2} \sumN
\overline{g}^{\mu \nu} \pa{\mu} \phi^i \pa{\nu} \phi^i
\right]
\label{CGHS action}
\ee
turns out to be a special case of~(\ref{our action}) with $\alpha = 0$
and $\beta = 1$ if we define
$\phi = -\e{-2\Phi}$ and $g_{\mu \nu} = \e{-2\Phi} \overline{g}_{\mu \nu}$.
On the other hand,
the case $\alpha \neq 0$ corresponds to the usual
two dimensional gravity theory coupled to scalar fields,
one of which has a background charge.
So,
we will call the theory~(\ref{our action}) with $\alpha = 0$
the dilaton gravity theory
and the one with $\alpha \neq 0$ the $\slr$ gravity theory.
The reason why we have used the term ``$\slr$''
will become clear in the following sections.
After we quantize the theory,
we will calculate correlation functions for ``tachyons''\footnote{
As will be explained,
``tachyons'' in this paper may not correspond to tachyons
in the conformal gauge theory.}.
It turns out that the screening charges of the symmetry algebra
play the role of the cosmological constant terms
in the conformal gauge when one computes correlation functions.
\section{Classical Theory}
\label{Classical Theory}
The action~(\ref{our action}) can be written as
\ba
\lefteqn{S = \int_{\Sigma} d^2 x (-det(g))^{- \frac{1}{2}}
\left[
  \frac{\alpha}{2}
  \left\{
    g_{11} \left( \pa{0} \phi \right)^2
    + g_{00} \left( \pa{1} \phi \right)^2
    - 2 g_{01} \pa{0} \phi \pa{1} \phi
  \right\}
  + \Lambda_0 \; det(g)
\right. }
\nonu
\A\A
\left.
  + \beta
  \left\{
    \pa{0} \phi \pa{0} g_{11}
    + \pa{1} \phi \pa{1} g_{00}
    - 2 \pa{0} \phi \pa{1} g_{01}
    + \frac{g_{01}}{g_{11}}
    \left(
      \pa{0} \phi \pa{1} g_{11}
      - \pa{1} \phi \pa{0} g_{11}
    \right)
  \right\}
\right.
\nonu
\A\A
\left.
  + \frac{1}{2} \sum_{i=1}^{N} \left\{
    g_{11} \left( \pa{0} \phi^i \right)^2
    + g_{00} \left( \pa{1} \phi^i \right)^2
    - 2 g_{01} \pa{0} \phi^i \pa{1} \phi^i
  \right\}
\right]
+ (\mbox{surface term}) .
\label{action expressed by Lagrangian}
\ea
If we assume $x^0$ is the ``time'',
the Lagrangian does not contain
``velocities'' of the $g_{00}$ and $g_{01}$ components.
Therefore they are Lagrange multipliers.
This fact means that the canonical momenta for $g_{00}$ and $g_{01}$
become zero,
reflecting the gauge invariance of the theory.
After the gauge is fixed,
the Hamiltonian becomes as follows:
\be
H = \oint dx^1 \left(
\frac{\sqrt{-det(g)}}{g_{11}} G_0 + \frac{g_{01}}{g_{11}} G_1
\right) + (\mbox{surface term}),
\label{Hamiltonian}
\ee
where
\ba
G_0 \A = \A \frac{\alpha}{2} \left( \pa{1} \phi \right)^2
  - \frac{\alpha}{2 \beta^2} \left( g_{11} p^{11} \right)^2
  + \frac{1}{\beta} g_{11} p^{11} \pi
  + \beta \frac{\pa{1} g_{11}}{g_{11}} \pa{1} \phi
  - 2 \beta \pa{1}^2 \phi + g_{11} \Lambda_0
\nonu
\A\A
  + \frac{1}{2} \sum_{i=1}^{N}
  \left\{
    \left( \pi_i \right)^2 + \left( \pa{1} \phi^i \right)^2
  \right\} ,
\\
G_1 \A = \A \pi \pa{1} \phi - p^{11} \pa{1} g_{11} - 2 g_{11} \pa{1} p^{11}
  + \sum_{i=1}^{N} \pi_i \pa{1} \phi^i .
\label{G0 G1}
\ea
It is easy to see that $G_0$ and $G_1$ are secondary constraints
and there are no more constraints.
Note that since the Hamiltonian is a linear combination of constraints,
all physical observables do not evolve as time goes by.
The two secondary constraints obtained above generate diffeomorphism
transformations.
In fact,
$T_{\pm} = \frac{1}{2} \left( G_0 \pm G_1 \right)$ obey Virasoro algebras
\be
\Pb{T_{\pm} (x^1 )}{T_{\pm} ({x^1}')} = \pm
\left( T_{\pm} (x^1 ) + T_{\pm} ({x^1}') \right)
\pa{1} \delta ( x^1 - {x^1}' ).
\label{two Virasoro algebra}
\ee
The equation~(\ref{two Virasoro algebra}) also means that $G_0$ and $G_1$
are first class constraints.
\par
We introduce new phase space variables which are
very useful when we quantize the theory.
The change of variables is divided in two steps.
First,
let us define
\be
\chi \equiv 4 \beta^2 g_{11}, \quad
\omega \equiv \frac{1}{4 \beta^2}
  \left(
    p^{11} - \beta \frac{\pa{1} \phi}{g_{11}}
  \right),
\quad
P_{S}^0 \equiv \pi + \beta \frac{\pa{1} g_{11}}{g_{11}}, \quad
P_{S}^1 \equiv \pa{1} \phi,
\label{chi omega PS0 PS1}
\ee
\be
P^{\pm}_i \equiv \frac{1}{\sqrt{2}} \left( \pi_i \pm \pa{1} \phi^i \right)
\iN .
\label{Pi+-}
\ee
Secondly,
following \cite{AM},
we define the following fields,
\ba
J^+ \A \equiv \A \frac{1}{2g_{11}} \left(G_1 -G_0 \right) + \frac{\Lambda_0}{2}
= \alpha \chi \omega^2
  - 2 \beta \omega \left( P_{S}^0 - \alpha P_{S}^1 \right)
  - 4 \beta^2 \pa{1} \omega
  - \frac{2 \beta^2}{\chi} \sumN \left( P_{i}^- \right)^2 ,
\\
J^0 \A \equiv \A \chi \omega + \gamma \left( P_{S}^0 + \alpha P_{S}^1 \right)
  + 2 \beta P_{S}^1 ,
\qquad
J^- \equiv \chi ,
\qquad
P_{S} \equiv P_{S}^0 + \alpha P_{S}^1 ,
\label{J+ J0 J- PS}
\ea
where we have introduced a free parameter $\gamma$ for later use.
The most remarkable thing about the above definition is that it resembles
the Wakimoto construction for the $\slr$ algebra \cite{Wakimoto}
except for the last term of $J^+$.
But it turns out that this difference does not matter.
In fact,
the Poisson brackets of these variables are
\ba
\Pb{ J^0 ( x^1 )}{ J^{\pm} ({x^1}')}
\A = \A \pm J^{\pm} ( x^1 ) \delta (x^1 - {x^1}') ,
\label{Pb of J0 Jpm}
\\
\Pb{ J^+ ( x^1 )}{J^- ({x^1}')} \A = \A
  -2 \left( \alpha J^0 (x^1 ) - (\beta + \alpha \gamma) P_S (x^1 ) \right)
  \delta (x^1 - {x^1}')
\nonu
\A\A
  + 4 \beta^2 \pa{1} \delta (x^1 - {x^1}') ,
\\
\Pb{ J^0 ( x^1 )}{J^0 ({x^1}')}
\A = \A 2 \gamma (2 \beta + \alpha \gamma) \pa{1} \delta (x^1 - {x^1}') ,
\\
\Pb{ J^0 ( x^1 )}{P_S ({x^1}')}
\A = \A 2 (\beta + \alpha \gamma) \pa{1} \delta (x^1 - {x^1}') ,
\\
\Pb{ P_S ( x^1 )}{P_S ({x^1}')}
\A = \A 2 \alpha \pa{1} \delta (x^1 - {x^1}') ,
\label{Pb of PS PS}
\ea
and the other brackets are zero.
In the case $\alpha \neq 0$,
we can put $\gamma = \frac{-\beta}{\alpha}$
and~(\ref{Pb of J0 Jpm}) --~(\ref{Pb of PS PS}) become
the commutation relations for the $\slr \otimes U(1)$ Kac-Moody algebra.
On the other hand,
if $\alpha = 0$,
then those brackets form
the Kac-Moody version of the centrally extended Poincar\'{e} algebra
\cite{Jackiw}.
In this case the simplest choice for $\gamma$ is $\gamma = 0$.
We call these current-like new phase space variables $\slr$
variables for brevity.
In the light-cone gauge,
which is defined by the gauge fixing condition $g_{00}=0$ and $g_{01}=-1$,
it turns out that the $\slr$ variables are the generators
of the residual symmetry \cite{TU}.
\par
As for the Poisson brackets of $\slr$ variables and matter ones,
all brackets are zero except for
\be
\Pb{J^+ ( x^1 )}{P^{-}_i ( {x^1}' )} =
4\beta^2 \frac{P^{-}_i ( x^1 )}{J^- ( x^1 )} \pa{1} \delta (x^1 - {x^1}') \iN .
\label{Pb of J+ P-}
\ee
This fact causes difficulties when we quantize the theory
because the inverse of the operator $J^-$ must be taken into account.
However,
if we put the non-zero modes of $P^{-}_i$ to be zero,
the Dirac bracket for $J^+$ and $P^{-}_i$ vanishes.
In this sense we will deal with ``chiral'' matter in this paper.
\par
Before embarking on the quantization of the theory,
we must construct the BRST charge.
To do that we will employ
the so-called BFV methods (\cite{Henneaux} is a good review for this theory).
According to them the BRST charge is constructed from the first class
constraints and their structure constants.
Since we obtained two first class constraints,
$G_0$ and $G_1$,
it is natural to use them.
However,
we can also choose linear combinations of them
which have forms easy to deal with,
namely,
\be
J^+ - \frac{\Lambda_0}{2}
= \frac{1}{2 g_{11}} \left( G_1 - G_0 \right) \approx 0 ,
\qquad
T_{G} + T_{M}^+ = G_1 \approx 0 ,
\label{constraints}
\ee
where
\ba
T_{G} \A = \A \frac{1}{4 \beta^2}
  \left(
    J^- J^+ - \alpha \left( J^0 \right)^2
  \right)
  - \pa{1} J^0
  + \frac{\beta + \alpha \gamma}{2 \beta^2} J^0 P_S
  - \frac{\gamma (2 \beta + \alpha \gamma)}{4 \beta^2} ( P_S )^2
  + \gamma \pa{1} P_S ,
\label{TG}
\\
T_{M}^+ \A = \A \frac{1}{2} \sumN (P^{+}_i)^2 .
\label{TM+}
\ea
$T_G$ is nothing but the Sugawara form
for the $\slr \otimes U(1)$ Kac-Moody algebra (in the case $\alpha \neq 0$)
or the Kac-Moody version of the centrally extended Poincar\'{e}
algebra (in the case $\alpha = 0$) with twisted terms,
namely derivative terms for $J^0 $ and $P_S $.
The BRST charge constructed from~(\ref{constraints}) by BFV method is
\be
Q = \oint dx^1 \, \left\{
    c^+ \left( J^+ - \frac{\Lambda_0}{2} \right)
    + c \left( T_G +T_{M}^+ \right) + c c^+ \, \pa{1} b^+ - b c \pa{1} c
  \right\} ,
\label{classical BRST charge}
\ee
where $b$,
$c$,
$b^+$ and $c^+$ are fermionic ghost fields.
\section{Quantization}
\label{Quantization}
In this section we will quantize the system
respecting the diffeomorphism symmetry of the classical theory.
Since the ``space'' is a circle,
we can decompose the fields into Fourier components,
\ba
J^{\pm ,0} (x^1 ) \A = \A \frac{1}{2 \pi} \sumz{n} J_{n}^{\pm ,0}
\e{-\iu n x^1},
\qquad
P_S (x^1 ) = \frac{\zeta}{2 \pi} \sumz{n} a_{n}^S \e{-\iu n x^1} ,
\label{mode expansion of J PS}
\\
P^{+}_i (x^1 ) \A = \A \frac{1}{\sqrt{2 \pi}} \sum_{n \in {\bf Z}}
a^{i}_{n} \e{- \iu n x^1} \iN ,
\\
b (x^1 ) \A = \A \frac{\iu}{2 \pi} \sumz{n} b_n \e{- \iu n x^1},
\qquad
c (x^1 ) = \sumz{n} c_n \e{- \iu n x^1} ,
\\
b^+ (x^1 ) \A = \A \frac{\iu}{2 \pi} \sumz{n} b_{n}^+ \e{- \iu n x^1},
\qquad
c^+ (x^1 ) = \sumz{n} c_{n}^+ \e{- \iu n x^1} ,
\\
T_{G} (x^1 ) \A = \A \frac{1}{2 \pi} \sumz{n} L_{n}^{G} \e{-\iu n x^1} ,
\qquad
T_{M}^+ (x^1 ) = \frac{1}{2 \pi} \sumz{n} L^{M}_n \e{- \iu n x^1} ,
\ea
where we have introduced an undetermined normalization constant $\zeta$
and
\be
L^{M}_m = \frac{1}{2} \sumN \sumz{n} \nm{a_{m-n}^i a_{n}^i}
\ee
are the Virasoro generators for the matter sector
with the central charge $c_M = N$.
The normal ordering symbol($:$ $:$) means that non-negative modes are
to the right of negative ones with respect to its arguments.
Since real variables in the classical theory become
Hermitian operators in the quantum theory,
the Hermiticity relations turn out to be that
$\hc{J^{\pm ,0}_m} = J^{\pm ,0}_{-m}$,
$\hc{a^{i}_m} = a^{i}_{-m}$,
$\hc{b_m} = b_{-m}$,
$\hc{c_m} = c_{-m}$,
$\hc{b^{+}_m} = b^{+}_{-m}$,
$\hc{c^{+}_m} = c^{+}_{-m}$,
$\hc{L^{G}_{m}} = L^{G}_{-m}$ and $\hc{L^{M}_{m}} = L^{M}_{-m}$,
while the Hermiticity relations for $a^{S}_m$ depend upon $\zeta$.
The conventional quantization procedure requires
that Poisson brackets must be replaced by commutators multiplied by $-\iu$.
We apply this procedure for $P^{+}_i$,
$b$,
$c$,
$b^+$ and $c^+$.
\ba
\com{ a^{i}_m }{ a^{j}_n } \A = \A m \delta^{ij} \delta_{m+n,0} \ijN
\\
\acom{ b_m }{ c_n } \A = \A
\acom{ c_m }{ b_n } =
\acom{ b_{m}^+ }{ c_{n}^+ } =
\acom{ c_{m}^+ }{ b_{n}^+ } = \delta_{m+n,0} .
\ea
As for the Poisson bracket algebra~(\ref{Pb of J0 Jpm}) --~(\ref{Pb of PS PS}),
we assume that these algebraic relations are preserved
also in the quantum theory except for the possible renormalization
of $c$-number ``anomaly'' terms,
namely,
\ba
\com{ J_{m}^0 }{ J_{n}^{\pm} } \A = \A
\pm \iu J_{m+n}^{\pm} ,
\label{commutation relation of J0 J+-}
\\
\com{ J_{m}^+ }{ J_{n}^- } \A = \A
  -2 \iu \left(
    \alpha J_{m+n}^0 - (\beta + \alpha \gamma) \zeta a_{m+n}^S
  \right)
  -2 \left(
    \kappa \alpha - \kappa' (\beta + \alpha \gamma)
  \right) m \delta_{m+n,0} ,
\label{commutation relation of J+ J-}
\\
\com{ J_{m}^0 }{ J_{n}^0 } \A = \A \kappa m \delta_{m+n,0} ,
\qquad
\com{ J_{m}^0 }{ a_{n}^S } = \frac{\kappa'}{\zeta} m \delta_{m+n,0} ,
\qquad
\com{ a_{m}^S }{ a_{n}^S }
= \frac{4 \pi \alpha}{\zeta^2} m \delta_{m+n,0}
\label{commutation relation of aS aS}
\ea
with
\be
4 \pi \alpha (\beta + \alpha \gamma ) = \alpha \kappa' ,
\label{condition for anomaly}
\ee
where $\kappa$ and $\kappa'$ are unknown parameters.
The equation~(\ref{condition for anomaly}) is one of the conditions
to make Jacobi identities hold.
We will study the case $\alpha \neq 0$ and the case $\alpha = 0$ separately.
\par
In the case where $\alpha \neq 0$,
if we put $\gamma = -\frac{\beta}{\alpha}$ and $\zeta = \sqrt{4 \pi \alpha}$,
then $\kappa' = 0$ from~(\ref{condition for anomaly}).
Consequently,
the non-trivial commutation relations become
\ba
\com{ J_{m}^0 }{ J_{n}^{\pm} } \A = \A
  \pm \iu J_{m+n}^{\pm} ,
\qquad
\com{ J_{m}^+ }{ J_{n}^- } =
  -2 \iu J_{m+n}^0 + k m \delta_{m+n,0} ,
\\
\com{ J_{m}^0 }{ J_{n}^0 } \A = \A - \frac{k}{2} m \delta_{m+n,0} ,
\qquad
\com{ a_{m}^S }{ a_{n}^S } = m \delta_{m+n,0} ,
\ea
where $k=-2 \kappa$ and $J_{m}^-$ is replaced by $\alpha J_{m}^- $.
This algebra is the $\slr \otimes U(1)$ Kac-Moody algebra.
Next,
we will construct $T_G$ from $\slr$ variables as in~(\ref{TG}).
Recall that classical $T_G$ is the generator of the spatial diffeomorphism
in the gravity sector.
Since we respect the diffeomorphism symmetry,
Poisson brackets for $T_G$ and the $\slr$ variables
are replaced by commutators with the same structure constants
taking into account some possible $c$-number renormalizations.
We know that the Fourier components of the required $T_G$ become
\ba
L_{m}^{G} \A = \A
\frac{1}{k+2} \sumz{n} \left(
  \frac{1}{2} \nm{J_{m-n}^- J_{n}^+} +  \frac{1}{2} \nm{J_{m-n}^+ J_{n}^-}
  - \nm{J_{m-n}^0 J_{n}^0}
\right) + \iu m J_{m}^0
\nonu
\A\A
  + \frac{1}{2} \sumz{n} \nm{a_{m-n}^S a_{n}^S} + \iu m Q_S a_{m}^S
  + \left(
    \frac{{Q_S}^2}{2} - \frac{k}{4}
  \right) \delta_{m,0} ,
\label{alpha neq 0 Sugawara form}
\ea
where $Q_S = \frac{\beta}{\sqrt{\pi \alpha}}$.
It is a well-known fact that the generators for the $\slr$ Kac-Moody algebra
can be expressed by free fields \cite{Wakimoto} such that
\ba
J^{+}_m \A = \A \beta_m ,
\qquad
J^{0}_m = - \iu \sumz{n} \nm{\beta_{m-n} \gamma_n}
+ \iu \xi a^{C}_m - \frac{\iu}{2} \delta_{m,0} ,
\label{alpha neq 0 free field representation of J+ J0}
\\
J^{-}_m \A = \A - \sumz{n,l} \nm{\beta_{m-n-l} \gamma_{n} \gamma_{l}}
+ 2 \xi \sumz{n} \nm{\gamma_{m-n} a^{C}_n}
- (1 + k m ) \gamma_m ,
\label{alpha neq 0 free field representation of J-}
\ea
where $\xi^2 = 1 + \frac{k}{2}$
and the canonical commutation relations for the $\beta$,
$\gamma$ and $a^C$ fields are $\com{ \beta_m }{ \gamma_n } = \delta_{m+n,0}$
and $\com{ a^{C}_m }{ a^{C}_n } = m \delta_{m+n,0}$.
The normal ordering for $\beta$ and $\gamma$ zero modes is defined such that
$\nm{\beta_0 \gamma_0} = \gamma_0 \beta_0$
and the Hermiticity relations become $\hc{\beta_m} = \beta_{-m}$,
$\hc{\gamma_m} = - \gamma_{-m}$ and $\hc{a^{C}_m} = -a^{C}_{-m}$ ($a^{C}_{-m}$)
if $k+2>0$ ($k+2<0$).
\par
In the case where $\alpha = 0$,
if we put $\zeta = \kappa'$ and redefine
$J_{m}^0 \rightarrow J_{m}^0 + \frac{\zeta \kappa}{2 \kappa'} a_{m}^S$,
then~(\ref{commutation relation of J0 J+-})
--~(\ref{commutation relation of aS aS}) become
\ba
\com{ J_{m}^0 }{ J_{n}^{\pm} } \A = \A
  \pm \iu J_{m+n}^{\pm} ,
\qquad
\com{ J_{m}^+ }{ J_{n}^- } =  \iu a_{m+n}^S + m \delta_{m+n,0} ,
\\
\com{ J_{m}^0 }{ a_{n}^S } \A = \A m \delta_{m+n,0} ,
\ea
where we have replaced $J_{m}^-$ by $2 \beta \kappa' J_{m}^- $.
This algebra is the Kac-Moody version of the centrally extended Poincar\'{e}
algebra.
Next,
we must determine quantum $T_G$ expressed by $\slr$ variables.
In the case of the $\slr$ gravity,
$T_G$ is nothing but the Sugawara form for the $\slr$ Kac-Moody algebra
with twisted term.
In general,
for a Kac-Moody algebra associated with a semi-simple Lie algebra,
the Sugawara form can be constructed
if one knows the Killing form of the Lie algebra.
In the present case,
the associated Lie algebra is the centrally extended Poincar\'{e}
algebra which is not semi-simple and its Killing form is degenerate.
However,
fortunately there is a bilinear form on the centrally extended Poincar\'{e}
algebra which has the same property as the Killing form
and is not degenerate \cite{Jackiw}.
Using this bilinear form we can construct $T_G$ whose Fourier components are
\ba
L_{m}^{G} \A = \A
\sumz{n} \left(
  \frac{1}{2} \nm{J_{m-n}^- J_{n}^+} +  \frac{1}{2} \nm{J_{m-n}^+ J_{n}^-}
  + \nm{J_{m-n}^0 a_{n}^S}
\right) + \iu m J_{m}^0
\nonu
\A\A
  + \frac{1}{2} \sumz{n} \nm{a_{m-n}^S a_{n}^S} + \iu m Q_S a_{m}^S
  + Q_S \delta_{m,0} ,
\label{alpha=0 Sugawara form}
\ea
where $Q_S$ is a real free parameter.
There exists a free field representation for $\slr$ variables
also in the case of the dilaton gravity theory,
namely,
\ba
J^{+}_m \A = \A \beta_m ,
\qquad
J^{0}_m = - \iu \sumz{n} \nm{\beta_{m-n} \gamma_n}
+ a^{C}_m - \frac{1}{2} a^{S}_m - \frac{\iu}{2} \delta_{m,0} ,
\label{alpha=0 free field representation of J+}
\\
J^{-}_m \A = \A \iu \sumz{n} \nm{\gamma_{m-n} a^{S}_n} - m \gamma_m ,
\label{alpha=0 free field representation of J-}
\ea
with canonical commutation relations,
$\com{ \beta_m }{ \gamma_n } = \delta_{m+n,0}$ and
$\com{ a^{C}_m }{ a^{S}_n } = m \delta_{m+n,0}$.
\par
$L^{G}_m$ expressed by free fields has the same form
in both the $\slr$ gravity theory and the dilaton gravity theory,
namely,
\ba
\lefteqn{
L^{G}_m = \sumz{n} (m+n) \nm{\gamma_{-n} \beta_{m+n}}
}
\nonu
\A\A
+ \frac{1}{2} \sumz{n} \nm{a^{D}_{m-n} a^{D}_n} + \iu ( m+1 ) Q_D a^{D}_m
+ \frac{1}{2} \sumz{n} \nm{a^{L}_{m-n} a^{L}_n} + ( m+1 ) Q_L a^{L}_m ,
\ea
where
\ba
Q_D \A \equiv \A \case{Q_S}{$\slr$ Gravity}
{\frac{1}{\sqrt{2}} \left( Q_S + 1 \right)}{Dilaton Gravity} ,
\\
Q_L \A \equiv \A \case{\frac{1}{2 \xi}-\xi}{$\slr$ Gravity}
{\frac{1}{\sqrt{2}} \left( Q_S - 1 \right)}{Dilaton Gravity} ,
\\
a^{D}_m \A \equiv \A \case{a^{S}_m - \iu Q_D \delta_{m,0}}{$\slr$ Gravity}
{\frac{1}{\sqrt{2}} \left( a^{S}_m + a^{C}_m \right) - \iu Q_D \delta_{m,0}}
{Dilaton Gravity} ,
\\
a^{L}_m \A \equiv \A \case{a^{C}_m - Q_L \delta_{m,0}}{$\slr$ Gravity}
{\frac{\iu}{\sqrt{2}} \left( a^{S}_m - a^{C}_m \right) - Q_L \delta_{m,0}}
{Dilaton Gravity} .
\ea
The condition $Q_D - Q_L = \sqrt{2}$ distinguishes the dilaton gravity theory.
\par
The quantum BRST charge $Q$ and the total Virasoro generator $L_m$
are defined as follows:
\ba
Q \A = \A
\sumz{n} \nm{c^{+}_{-n} \left( \beta_{n} - \Lambda \delta_{n,0} \right)}
+ \sumz{n} \nm{c_{-n} \left(
L^{G}_n + L^{M}_n + L^{gh+}_n
\right)}
\nonu
\A\A
- \frac{1}{2} \sumz{m,n} (m-n) \nm{c_{-m} c_{-n} b_{m+n}} - c_0 ,
\label{quantum BRST charge}
\\
L_m \A \equiv \A \acom{ Q }{ b_m }
= L^{G}_m + L^{M}_m + L^{gh+}_m + L^{gh}_m ,
\ea
where
\be
L_{m}^{gh+} = \sumz{n} (m+n) \nm{c_{-n}^+ b_{m+n}^+} ,
\qquad
L_{m}^{gh} = \sumz{n} (m-n) \nm{b_{m+n} c_{-n}} - \delta_{m,0}
\ee
are the Virasoro generators for $b^+$$c^+$ and $b$$c$ ghosts respectively
with the central charges $c_{gh+} = -2$ and $c_{gh} = -26$.
The normal ordering for $b^+$$c^+$ and $b$$c$ ghost zero modes is defined such
that $\nm{b^{+}_0 c^{+}_0} = - c^{+}_0 b^{+}_0$ and $\nm{b_0 c_0} = - c_0 b_0$.
In the above definition for
the quantum BRST charge~(\ref{quantum BRST charge}),
we replaced $\pi \Lambda_0$ by $\Lambda$ taking into account
a possible renormalization for the cosmological constant.
It is easy to check that
the BRST charge defined by~(\ref{quantum BRST charge}) is nilpotent
if and only if the total central charge vanishes,
namely,
\be
Q_{D}^2 - Q_{L}^2 = 2 - \frac{N}{12}.
\label{constraint for QD QL}
\ee
\section{Correlation Functions}
\label{Correlation Functions}
An observable is an Hermitian operator
whose commutator with the BRST charge vanishes.
Here,
we will give some examples of observables.
First,
we introduce a vertex operator which is defined such that
\be
\tvo{}{} \equiv \nm{\e{\iu \vec{p} \cdot
( \vec{\phi} ( x^1 ) + \vec{p}_Q x^1 )}}
= \e{\iu \vec{p} \cdot \vec{\phi}_- (x^1)}
\e{\iu \vec{p} \cdot \vec{\phi}_+ (x^1)}
\e{\iu \vec{p} \cdot \left( \vec{q} + (\vec{a}_0 + \vec{p}_Q) x^1 \right)},
\label{vo}
\ee
where
$\vec{p}_Q = \left( p^{L}_{Q} , p^{D}_{Q} , p^{1}_{Q}
, \ldots , p^{N}_Q \right)
\equiv \left( Q_L , \iu Q_D , 0 , \ldots ,0 \right)$
and the components of $\vec{\phi} (x^1)$ are
\be
\phi^I (x^1) \equiv q_I + a^{I}_0 x^1
+ \iu \sumn \frac{a^{I}_n}{n} \e{-\iu n x^1}
\qquad
(I=D,L,1,\ldots,N) .
\label{def phiI}
\ee
In equation~(\ref{vo}) we have defined that
$\phi^{I}_+ ( x^1 ) (\phi^{I}_- ( x^1 ))$
is the positive(negative) frequency mode part of $\phi^I ( x^1 )$,
and $q_D$,
$q_L$ and $q_i$ are canonical conjugate operators for $a^{D}_0$,
$a^{L}_0$ and $a^{i}_0$ respectively,
which satisfy $\com{q_I}{a^{J}_0} = \iu \delta^{J}_I$.
Note that $\tvo{}{}$ is Hermitian when the exponents of
the normal ordered exponential functions are Hermitian.
Therefore $\tvo{}{}$ is a candidate for an observable.
To see the condition that $\tvo{}{}$,
or something made from $\tvo{}{}$,
becomes an observable,
we must calculate the commutator of $\tvo{}{}$ with the BRST charge,
which becomes
\be
\com{Q}{\tvo{}{}} = -\iu \pa{1} \w{}{}
- \iu \left( h (\vec{p}) -1 \right) \pa{1} c (x^1) \tvo{}{} ,
\label{com Q tvo}
\ee
where $\w{}{} \equiv c(x^1) \tvo{}{}$ and
$h (\vec{p}) \equiv \frac{1}{2} \vec{p} \cdot \vec{p}
+ \vec{p} \cdot \vec{p}_Q$ is the conformal weight of $\tvo{}{}$
with respect to the total Virasoro algebra.
Therefore if $h (\vec{p}) = 1$,
the integral of $\tvo{}{}$ with respect to $x^1$ is an observable
if $\tvo{}{}$ takes the same values both at the beginning and the end point
of the integral contour.
We will call $\tvo{}{}$ with the conformal weight $1$ a tachyon vertex
operator\footnote{
As we shall see in the following,
$\tvo{}{}$ is {\it not} a tachyon vertex operator in the Liouville theory,
although it looks like a chiral component of one.
But we will use this term for simplicity.}.
Under the condition $h (\vec{p}) = 1$,
$\w{}{}$ also becomes an observable.
The following operator is also an observable.
\be
\psi (x^1) = \beta (x^1) \tvo{\psi}{} ,
\ee
where $\beta(x^1 ) = \frac{1}{2\pi} \sumz{n} \beta_n \e{-\iu n x^1}$ and
\be
\vec{p}_{\psi} = \left( p^{L}_{\psi} , p^{D}_{\psi} , p^{1}_{\psi}
, \ldots , p^{N}_{\psi} \right)
\equiv \case{\left( -\frac{1}{\xi} , 0 , 0 , \ldots ,0 \right)}
{$\slr$ Gravity}
{\left( - \frac{1}{\sqrt{2}} , - \frac{\iu}{\sqrt{2}} , 0 , \ldots ,0 \right)}
{Dilaton Gravity} .
\ee
As is easily checked,
the commutators of $\psi (x^1)$ and the generators
of the $\slr$ Kac-Moody algebra
or the Kac-Moody version of the centrally extended Poincar\'{e} algebra
become zero or total derivatives.
Therefore $\psi (x^1)$ can be considered as a screening current
of these algebras \cite{BFO}
and the integral of $\psi (x^1)$ is an observable.
Note that the forms of screening currents are similar to the integrands of
the cosmological constant terms in both the Liouville theory \cite{DDK}
and the dilaton gravity theory \cite{MSTU}.
In fact,
as will be shown in the following,
the integrals of these screening currents
play the role of cosmological constant terms in correlation functions
\cite{FK}.
\par
The $r+3$ point correlation function for tachyons are defined as follows:
\ba
\lefteqn{
Z \left( \vec{p}_a,\vec{p}_b,\vec{p}_c,\vec{p}_1,\ldots ,\vec{p}_r;s \right)
= \bra{0;\slc} b^{+}_0 \delta ( \beta_0 - \Lambda ) \w{a}{a}
}
\nonu
\A\A
\times
\w{b}{b} \w{c}{c}
\left[ \prod_{j=1}^{r} \int_{C_j} d x^{1}_j \tvo{j}{j} \right]
\left[ \prod_{k=1}^{s} \int_{S_k} d x^{1}_k \psi (x^{1}_k) \right]
\ket{0;\slc} ,
\label{correl}
\ea
where $\ket{0;\slc}$ is called the $\slc$ vacuum
defined by the following conditions,
\be
a^{I}_m \ket{0;\slc} = 0
\qquad ( I = D, L, 1, \ldots, N )
\quad (m \geq  0),
\label{def slc vac a}
\ee
\be
\beta_m \ket{0;\slc} = b^{+}_m \ket{0;\slc} = 0
\quad (m \geq 1),
\label{def slc vac beta b+}
\ee
\be
\gamma_m \ket{0;\slc} = c^{+}_m \ket{0;\slc} = 0
\quad (m \geq 0),
\label{def slc vac gamma c+}
\ee
\be
b_m \ket{0;\slc} = 0 \quad (m \geq -1),
\qquad
c_m \ket{0;\slc} = 0 \quad (m \geq 2).
\label{def slc vac bc}
\ee
The name of the $\slc$ vacuum comes from the fact that this vacuum vanishes
upon operation of the Virasoro generators $L_{-1}$,
$L_{0}$ and $L_{1}$ which are the generators for M\"{o}bius transformations
in conformal field theories.
The reason why we bring up the $\slc$ vacuum is that it is a physical state;
namely,
is annihilated by the BRST charge.
Note that if $h(\vec{p})=1$,
$\ket{\vec{p}} = \oint d x^1 \w{}{}
b^{+}_0 \delta ( \beta_0 - \Lambda ) \ket{0;\slc}$
is also a physical state which we will call a Fock vacuum.
Since $\w{}{}$ is independent of its position in a correlation function
of observables,
$\bra{0;\slc} b^{+}_0 \delta ( \beta_0 - \Lambda ) \w{a}{a}$ in~(\ref{correl})
can be considered as the bra corresponding to the Fock vacuum
$\ket{\vec{p}_a}$.
We inserted $s$ screening charges and three $\w{}{}$'s
in equation~(\ref{correl}).
If the number of the inserted $\w{}{}$'s is not three,
the correlation function vanishes identically because of the contribution
from $bc$ ghosts\footnote{
In the path integral language,
these three $c$ ghosts are needed to cancel the Grassmann integrals
of $c$ ghost zero modes which correspond to conformal Killing vectors.}.
The correlation function defined above may become singular because
the operator product $\tvo{1}{1} \tvo{2}{2}$ may be singular at
$x^{1}_1 = x^{1}_2$.
Therefore we must regularize it.
We employ the analytic continuation for the regularization scheme;
in practice,
we analytically continue the coordinates of the operators
$z \equiv \e{\iu x^1}$ from the unit circle to the whole complex plane
such that the absolute values of the $z$'s become smaller from left to right;
for example,
$|z_j| > |z_k|$ for $j<k$,
where $z_j = \e{\iu x^{1}_j}$.
Under the regularization scheme stated above and the mass shell conditions,
$h(\vec{p}_a) = h(\vec{p}_b) = h(\vec{p}_c)
= h(\vec{p}_1) = \cdots = h(\vec{p}_r) = h(\vec{p}_{\psi}) = 1$,
we obtain the following form for the correlation function,
\be
Z \left( \vec{p}_a,\vec{p}_b,\vec{p}_c,\vec{p}_1,\ldots ,\vec{p}_r;s \right)
= \bra{0;\slc} c_0 c_{-1} \ket{\vec{v}}
A \left( \vec{p}_a,\vec{p}_b,\vec{p}_c,\vec{p}_1,\ldots ,\vec{p}_r ;s \right) ,
\label{corr2}
\ee
where $\vec{v} = \vec{p}_a + \vec{p}_b + \vec{p}_c
+ \sum_{j=1}^r \vec{p}_j + s \vec{p}_{\psi}$.
Note that $\bra{0;\slc} c_0 c_{-1} \ket{\vec{v}}
= \delta \left( \vec{v} + 2 \vec{p}_Q \right)$\linebreak
$\times \bra{0;\slc} c_0 c_{-1} \ket{-2 \vec{p}_Q}$,
where the delta function ensures
the momentum conservation and $\bra{0;\slc} c_0 c_{-1} \ket{-2 \vec{p}_Q}$
is considered as the vacuum to vacuum amplitude,
by which the correlation function must be divided.
The amplitude
$A \left( \vec{p}_a,\vec{p}_b,\vec{p}_c,\vec{p}_1,\ldots ,\vec{p}_r ;s \right)$
in equation~(\ref{corr2}) can be written in the following form,
\ba
\lefteqn{
A \left( \vec{p}_a,\vec{p}_b,\vec{p}_c,\vec{p}_1,\ldots ,\vec{p}_r ;s \right)
= \left( \frac{\Lambda}{2\pi} \right)^s (-\iu)^{r+s}
\prod_{j=1}^r \int_{C_j} d z_j \prod_{k=1}^s \int_{S_k} d w_k
}
\nonu
\A\A
\times
\left[ \prod_{j=1}^r \sel{z_j}{a}{j} \sel{(z_j - 1)}{b}{j} \right]
\left[ \prod_{k=1}^s \sel{w_k}{a}{\psi} \sel{(w_k - 1)}{b}{\psi} \right]
\nonu
\A\A
\times
\left[ \prod_{i<j}^r \sel{(z_i - z_j)}{i}{j} \right]
\left[ \prod_{j<k}^s \sel{(w_j - w_k)}{\psi}{\psi} \right]
\left[ \prod_{j=1}^r \prod_{k=1}^s \sel{(z_j - w_k)}{\psi}{j} \right] .
\label{amplitude}
\ea
The paths of integrations are divided into three groups.
The paths which belong to the first group
start from $1$ and end at $\infty$.
The paths which belong to the second group
start from $0$ and end at $1$.
The paths which belong to the third group
start from $\infty$ and end at $0$.
$C_1 , \ldots , C_{r_1}$ and $S_1 , \ldots , S_{l}$
belong to the first group while
$C_{r_1 + 1} , \ldots , C_{r_1 + r_2}$ and $S_{l+1} , \ldots , S_{l+m}$
belong to the second group and
$C_{r_1 + r_2 + 1} , \ldots , C_{r_1 + r_2 + r_3}$
and $S_{l+m+1} , \ldots , S_{l+m+n}$
belong to the third group,
where $r_1 + r_2 + r_3 = r$ and $l+m+n = s$.
Moreover,
in each of the three groups,
the $S_k$'s lie above the $C_j$'s
and $C_j$($S_j$) lies above $C_i$($S_i$) if $i<j$.
\par
When $r=0$,
we can perform the integral of~(\ref{amplitude}) using the formula
of \cite{DF}.
The result is as follows:
\be
A \left( \vec{p}_a,\vec{p}_b,\vec{p}_c;s \right)
= \left( \frac{\Lambda}{2 \pi \iu} \right)^s \e{\iu \pi \theta} \J{l}{m}{n},
\label{three point}
\ee
where $\e{\iu \pi \theta}$ is an irrelevant phase factor and
$\J{l}{m}{n}$ in~(\ref{three point}) is defined as follows:
\ba
\lefteqn{
\J{l}{m}{n} = \prod_{k=1}^s \frac{\G{1-\rho}}{\G{1-k\rho}}
}
\nonu
\A\A
\times
\prod_{k=1}^l \frac{1}{\G{-\alpha - (s-k) \rho}}
\prod_{k=1}^m \frac{1}{\G{-\gamma - (s-k) \rho}}
\prod_{k=1}^n \frac{1}{\G{-\beta - (s-k) \rho}}
\nonu
\A\A
\times
\prod_{k=1}^{m+n} \G{1 + \alpha + (k-1) \rho}
\prod_{k=1}^{n+l} \G{1 + \gamma + (k-1) \rho}
\prod_{k=1}^{l+m} \G{1 + \beta + (k-1) \rho} ,
\ea
where $\alpha = \vec{p}_{\psi} \cdot \vec{p}_a$,
$\beta = \vec{p}_{\psi} \cdot \vec{p}_b$,
$\rho = \frac{1}{2} \vec{p}_{\psi} \cdot \vec{p}_{\psi}$
and $\gamma = - 2 - \alpha - \beta - 2 ( s - 1 ) \rho
= \vec{p}_{\psi} \cdot \vec{p}_c$.
\par
As an immediate example of a correlation function,
we will calculate the three point function for the dilaton gravity theory.
In this case,
owing to the fact $\rho =\frac{1}{2} \vec{p}_{\psi} \cdot \vec{p}_{\psi} =0$,
the amplitude becomes much simpler;
\ba
A \left( \vec{p}_a,\vec{p}_b,\vec{p}_c;s \right)
\A = \A \e{\iu \pi \theta}
\left( \frac{\iu \Lambda}
{2 \G{-\vec{p}_a \cdot \vec{p}_{\psi}}
\G{-\vec{p}_b \cdot \vec{p}_{\psi}}
\G{-\vec{p}_c \cdot \vec{p}_{\psi}}} \right)^s
\nonu
\A\A
\times
\frac{1}
{\left( \sin \left( \pi \vec{p}_a \cdot \vec{p}_{\psi} \right) \right)^l
\left( \sin \left( \pi \vec{p}_c \cdot \vec{p}_{\psi} \right) \right)^m
\left( \sin \left( \pi \vec{p}_b \cdot \vec{p}_{\psi} \right) \right)^n } .
\ea
Compared with the results of \cite{MSTU},
our amplitude seems to be a chiral part of theirs.
\par
In the rest of this section,
we will study correlation functions without matter $({\it i.e.} N=0)$.
For this purpose,
we will introduce the notion of chirality.
The solutions of the mass shell condition $h(\vec{p}) = 1$ have simple forms
in this case;
namely,
$p^L = \pm \iu (p^D + \iu Q_D ) - Q_L$.
We will say that the momentum with a plus(minus) sign has positive(negative)
chirality and write it as $\vec{p}^+$($\vec{p}^-$).
And the chiralities of a set of momenta
$(\vec{p}_a ,\vec{p}_b ,\vec{p}_c ,\vec{p}_1 , \ldots ,\vec{p}_r)$
is denoted by,
for example,
$(-,+,+,+, \ldots ,+)$ if the chirality of the momenta
$\vec{p}_b$,
$\vec{p}_c$,
$\vec{p}_1 , \ldots ,\vec{p}_r$ are positive
and that of $\vec{p}_a$ is negative.
Since we already know the general three point function
for the dilaton gravity theory,
we will investigate three point functions for the $\slr$ gravity theory
without matter.
It is easy to show that the amplitudes with chirality
$(+,+,+)$ and $(-,-,-)$ vanish.
Therefore we will study the amplitudes with chirality $(+,+,-)$ and $(-,-,+)$.
The amplitudes with other chiralities are obtained from these amplitudes
owing to the symmetry of the amplitudes.
When the chirality is $(+,+,-)$,
$\alpha + \beta = -1 + (2-s) \rho$ and $\gamma = -1 - s\rho$.
Therefore $\J{l}{m}{n}$ becomes as follows:
\ba
\J{l}{m}{n} \A = \A
\frac{-\pi \left( \G{-\rho} \right)^s}
{\G{1+\frac{1}{2} \vec{p}_a \cdot \vec{p}_a}
\G{1+\frac{1}{2} \vec{p}_b \cdot \vec{p}_b}
\G{1+\frac{1}{2} \vec{p}_c \cdot \vec{p}_c}}
\nonu
\A\A
\times
\frac{1}{\sin \left[ \pi \left(
\frac{1}{2} \vec{p}_a \cdot \vec{p}_a + n \rho \right) \right]}
\prod_{k=1}^{m}
\frac{\sin ( \pi k \rho )}
{\sin \left[ \pi \left(
\frac{1}{2} \vec{p}_a \cdot \vec{p}_a + (k+n) \rho \right) \right]} .
\ea
When the chirality is $(-,-,+)$,
$\alpha + \beta = -2 + \rho (1-s)$ and $\gamma = (1-s) \rho$.
It turns out that,
if $m \neq 0$,
the amplitude vanishes.
Therefore we assume $m = 0$.
\be
\J{l}{0}{n} = \frac{(-1)^n \left( \G{-\rho} \right)^s}
{\G{1+ \frac{1}{2} \vec{p}_c \cdot \vec{p}_c}}
\prod_{k=1}^s \frac{1}{k-1-\frac{1}{2} \vec{p}_a \cdot \vec{p}_a} .
\ee
These amplitudes seem to be those of open string theories \cite{TY}.
\section{Conclusions}
\label{Conclusions}
In this paper we investigated a canonical formalism for the two dimensional
quantum gravity theory,
formally,
in a gauge independent way.
Our method is useful at least in the light-cone gauge.
The resulting theory seems to be a chiral part of the Liouville theory,
{\it i.e.} the conformal gauge theory,
or a non-critical open string theory
reflecting the asymmetry property of the light-cone gauge.
However,
to compare with the results of the conformal gauge theory,
it is necessary to investigate the correspondence between the two;
for example,
we must know much about the correspondence between the ``tachyon'' vertex
operators in our theory and the tachyon vertex operators
in the conformal gauge theory.
The second problem is that we discarded half of the dynamical degrees
of freedom for the matter sector.
The theory including all the degree of freedom must be studied.
In the presence of the degree of freedom,
which we discarded in this paper,
the constraints we took in this paper are not convenient.
We tried to use some other constraints,
but they do not seem to be suitable for quantization.
\section*{Acknowledgement}
I am grateful to Y.Matsumura,
N.Sakai and Y.Tanii.
This work is inspired by the collaboration with them.
Especially,
we thank N.Sakai.
Without his useful suggestions,
this work could not have been done.
I am also grateful to N.Ishibashi,
T.Mishima,
H.Mukaida,
A.Nakamichi,
H.Shirokura and T.Ueno for discussions.
And I am greatly grateful to the colleagues of Tokyo Institute of Technology
for their hospitality.
Finally,
I wish to thank N.A.McDougall for a careful reading of the manuscript.

\end{document}